\documentclass[12pt,a4paper]{article}
\usepackage{authblk}
\usepackage{indentfirst}
\usepackage{graphicx}
\usepackage{epsfig}

\usepackage{latexsym}
\usepackage{amsmath}
\usepackage{amssymb}
\usepackage{amsfonts}
\usepackage{mathrsfs}
\usepackage{pifont}

\usepackage{amsthm}
\usepackage{float}
\usepackage{subfigure}
\usepackage{color}
\usepackage{multicol}

\usepackage[colorlinks,linkcolor=blue,filecolor=black,citecolor=blue]{hyperref}

\newtheorem{theorem}{Theorem}[section]
\newtheorem{lemma}[theorem]{Lemma}

 \newtheorem{proposition}[theorem]{Proposition}

 \newtheorem{definition}[theorem]{Definition}

  \newcommand{\bC}{\boldsymbol{C}}

 \newcommand {\bx} {\boldsymbol{x}}
  \newcommand {\bff} {\boldsymbol{f}}
  \newcommand {\bmu} {\boldsymbol{\mu}}
    \newcommand {\bSigma} {\boldsymbol{\Sigma}}
     \newcommand {\bLambda} {\boldsymbol{\Lambda}}
      \newcommand {\bsigma} {\boldsymbol{\sigma}}  
  \newcommand {\btheta} {\boldsymbol{\theta}}
 \newcommand {\by} {\boldsymbol{y}}
  \newcommand {\bY} {\boldsymbol{Y}}
 \newcommand {\bA} {\boldsymbol{A}}
   \newcommand {\bbm} {\boldsymbol{m}}
 \newcommand {\bX} {\boldsymbol{X}}
 
  \newcommand {\bepsilon} {\boldsymbol{\epsilon}}
    \newcommand {\bphi} {\boldsymbol{\phi}}

 \newcommand {\bF} {{\bf F}}

 \def\bx{\boldsymbol{x}}

\newcommand{\mbf}{\boldsymbol}

\usepackage{ulem}
\title{Numerical Method for Parameter Inference of Nonlinear ODEs with Partial Observations }
\author[1,2]{Yu Chen}
\author[3,2]{Jin Cheng}
\author[4]{Arvind Gupta}
\author[5,1,4,6]{Huaxiong Huang\thanks{Corresponding author: hhuang@uic.edu.hk}}
\author[7]{Shixin Xu\thanks{Corresponding author: shixin.xu@dukekunshan.edu.cn}}

\affil[1]{\small Centre for Quantitative Analysis and Modeling (CQAM), The Fields Institute for Research in Mathematical Sciences, 222 College Street, Toronto, Ontario, Canada.}
\affil[2]{{\small School of Mathematics, Shanghai University of Finance and Economics, Shanghai, China}}
\affil[3]{{\small School of Mathematical Sciences, Fudan University, Shanghai, 200433, China}}
\affil[4]{\small Computer Science, University of Toronto, Toronto, Ontario, Canada.}
\affil[5]{\small Joint Mathematical Research Centre of Beijing Normal University and BNU-HKBU United International College, Zhuhai, China}
\affil[6]{\small Department of Mathematics and Statistics, York University, Toronto, Ontario, Canada.}
\affil[7]{\small Duke Kunshan University, 8 Duke Ave, Kunshan, Jiangsu, China.}

\date{}
\begin{document}
\maketitle

\section*{Abstract}

Parameter inference of dynamical systems is a challenging task faced by many researchers and practitioners across various fields. In many applications, it is common that only limited variables are  observable. In this paper, we propose a  method for parameter inference of a system of nonlinear coupled ODEs with partial observations. Our method  combines fast Gaussian process based gradient matching (FGPGM) and deterministic optimization algorithms. By using initial values obtained by  Bayesian steps with low sampling numbers, our deterministic optimization algorithm is both accurate and efficient.


Key words: Gaussian Process, Parameter inference, Nonlinear ODEs, Partial observations 

\section{Introduction}

Many problems in science and engineering
can be modelled by systems of  Ordinary differential equations (ODEs). It is often difficult or impossible to measure
some parameters of the systems directly.
Therefore, various methods have been developed to estimate parameters based on available data. Mathematically, such problems are classified as inverse problems which have been widely studied \cite{Anger1990,Aster2005,Tarantola2005,Li2005}.
 They can be also treated as parameter inference in statistics \cite{Ramsay2007,kaipio2006}.

For nonlinear ODEs, standard  statistical inference is  time consuming  as  numerical integration 
is needed  after each update of the parameters \cite{Calderhead,Macdonald2015}. Recently, gradient matching techniques have been proposed  to circumvent the high computational cost of numerical integration \cite{Ramsay2007,Dondelinger2013,Niu2016,Gorbach2017}. These techniques are  based on minimizing the difference between the values obtained by two different approaches.
This usually  
 involves a process consisting of  two steps: data interpolation and parameter adaptation. 
 Among them,   nonparametric Bayesian modelling with Gaussian processes is one of the promising approaches. 
  Calderhead et al. \cite{Calderhead} proposed an
adaptive gradient matching method based on a product-of-experts approach and a marginalization over the derivatives of the state variables, which was proposed by Calderhead et al. \cite{Calderhead} and extended by Dondelinger et al. \cite{Dondelinger2013}.
 Barber \& Wang \cite{Barber2014} proposed a GPODE method in which the state variables are marginalized. Macdonald et al. \cite{Macdonald2015} provided an interpretation of the above paradigms.  Wenk et al. \cite{Wenk2019} proposed a fast Gaussian process based gradient matching (FGPGM) algrithm with theoretical framework in systems of nonlinear ODEs. our new approach is more accurate, robust and efficient.

For many practical problems, the variables are only partially observable, or not at all times.
As a consequence, parameter inference  is more challenging,
even for a coupled system where the parameters are uniquely determined by data of partially observed data under certain initial conditions. 
It is not clear whether the gradient matching techniques  can be applied to the case when there are latent variables.
 The Markov Chain Monte Carlo algorithm has the ability to side-step the issue of parameter identifiability in many cases, but
 convergence remains a serious issue \cite{Ramsay2007}. Therefore, we need to pay attention to the feasibility, accuracy, robustness and computational cost of numerical computations for such problems.

In this work, we focus on the case of parameter inference with partially observable  data. The main idea is to treat the observable and nonobservale variables differently.  For observable variables, we use the same approach as proposed by Wenk et al. \cite{Wenk2019}. For non-observable variablies, they are obtained only by using ODEs. To circumvent the high computational cost of sampling in Bayesian approaches, we also combine FGPGM with least square optimization method. The remaining part of the paper is organized as follows. In Section 2 we give the numerical method to deal with parameter identification problems with partial observation. Numerical examples are presented in Section3. Some concluding remarks are given in Section 4.


\section{Algorithm}
The main strategy of FGPGM is to minimize the mismatch between the data and the ODE solutions in a maximum likelihood sense, making use of the property that Gaussian process is closed under differentiation. 

\begin{figure}[H] 
\begin{center}
\includegraphics[width=0.6\textwidth]{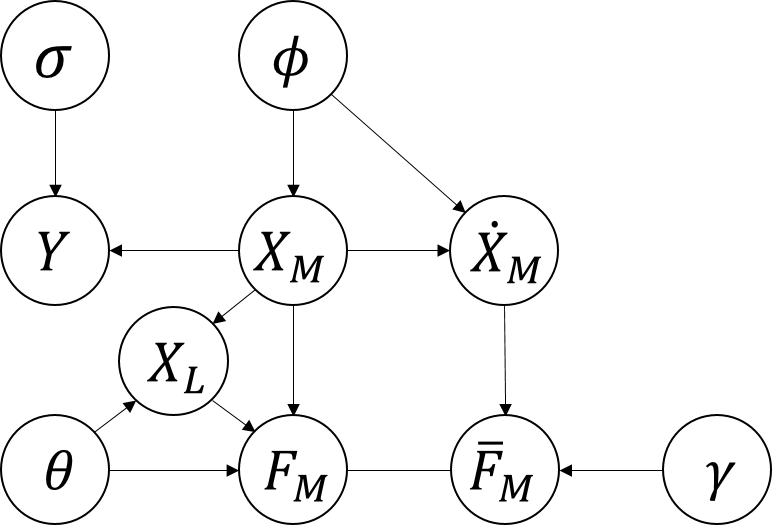}
\end{center}
  \caption{Probabilistic model with partial observable variables.}
\label{Numerical-harmonic-measure} 
\end{figure}

In this work, we would like to estimate the time-independent parameters $\mbf{\theta}$ of the following dynamical system  described by
\begin{equation}
    \dot{\mbf{X}}=\mbf{f}(\mbf{X},\mbf{\theta}).
\end{equation}
$\dot{\mbf{X}}$ is the vector of time derivative of the state $\mbf{X}$ and $\mbf{f}$ can be an nonlinear vector valued function.
We assume only parts of the variables are measurable and denote them as $\mbf{X}_M$. They are observed on discrete time points as $\mbf{Y}(t_i) (i=1,...N)$ with noise $\epsilon$ such that $\mbf{Y}=\mbf{X}_M+\epsilon$. We assume that the noise is Gaussian $\epsilon(t_i)\sim \mathcal{N}(\mbf{0},\sigma^2\mbf{I})$, then
\begin{equation}
\rho(\mbf{y}|\mbf{x}_M,\sigma)=\mathcal{N}(\mbf{y}|\mbf{x}_M,\sigma^2\mbf{I}),
\end{equation}
where $\mbf{x}_M$ and $\mbf{y}$ are the realizations of $\mbf{X}_M$ and $\mbf{Y}$ respectively.
The latent/unmeasurable variables are denoted as $\mbf{X}_L$, with $dim(\mbf{X}_M)+dim(\mbf{X}_L)=dim(\mbf{X})$. The idea of Gaussian process based gradient matching is as follows. Firstly, we put a Gaussian process prior on $\mbf{x}_M$,
\begin{equation}
\rho(\mbf{x}_M|\mbf{\mu}_M,\phi)=\mathcal{N}(\mbf{x}_M|\mbf{\mu}_M,\mbf{C}_{\phi}).
\end{equation}
Then according to Lemma \ref{conditionalmuandvariance} the conditional distribution of the $k$th state derivatives is 
\begin{equation}
\rho (\dot{\mbf{x}}_{M,k}|\mbf{x}_{M,k},\phi_{k})=\mathcal{N}(\dot{\mbf{x}}_{M,k}|\mbf{D}_k\mbf{x}_{M,k},\mbf{A}_k),
\end{equation}
where
\begin{equation}
\mbf{D}_k=\mbf{C}_{\phi_k}(\dot{\mbf{x}}_k,\mbf{x}_k)\mbf{C}_{\phi_k}(\mbf{x}_k,\mbf{x}_k)^{-1}(\mbf{x}_k-\mbf{\mu}_k),
\end{equation}
\begin{equation}
\mbf{A}_k=\mbf{C}_{\phi_k}(\dot{\mbf{x}}_k,\dot{\mbf{x}}_k)-\mbf{C}_{\phi_k}(\dot{\mbf{x}}_k,\mbf{x}_k)\mbf{C}_{\phi_k}(\mbf{x}_k,\mbf{x}_k)^{-1}\mbf{C}_{\phi_k}(\mbf{x}_k,\dot{\mbf{x}}_k).
\end{equation}
Here we have denoted $\mbf{C}_\phi$ as the covariance matrix. Its components are given by $\mbf{C}_\phi(i,j)=k_\phi(t_i,t_j)$ with respect to a kernel function $k_\phi$ parameterized by the hyperparameter $\phi$. For more details we refer to Appendix \ref{appendix-preliminaries}.
There is also a Gaussian noise with standard deviation $\gamma$ introduced to represent the model uncertainty 
\begin{equation}
\rho(\dot{\mbf{x}}|\mbf{x},\mbf{\theta},\gamma)=\mathcal{N}(\dot{\mbf{x}}|\mbf{f}(\mbf{x},\mbf{\theta}),\gamma\mbf{I}).
\end{equation}
We set up the following graphical probabilistic model to show the relationship between the variables (Fig. \ref{Numerical-harmonic-measure} ). Then the joint density can be represented by the following theorem.


\begin{theorem}
Given the modeling assumptions summarized in the graphical probabilistic model in Fig. \ref{Numerical-harmonic-measure}, 
 \begin{align}
 &\rho(\mbf{x}_M,\btheta|\mbf{y},\bphi,\bsigma,\mbf{\gamma})\notag\\
 &=\rho(\mbf{\theta})\mathcal{N}(\mbf{x}_M|\mbf{0},\mbf{C}_\phi)\mathcal{N}(\mbf{y}|\mbf{x}_M,\sigma^2\mbf{I})\mathcal{N}(\mbf{f}_M(\mbf{x}_M,\tilde{\mbf{x}}_L(\mbf{x}_M,\mbf{\theta}),\mbf{\theta}) |\mbf{D}\mbf{x}_M,\mbf{A}+\gamma\mbf{I}) \label{joint-density-proof}
 \end{align}
 where $\tilde{\mbf{x}}_L(\mbf{x}_M,\mbf{\theta})$ involved in $\mbf{f}_M$ is the solution determined by $\mbf{\theta}$ and $\mbf{x}_M$.
 \label{jointdensity-partial}
\end{theorem}
The proof can be found in Appendix \ref{appendix-proof}. 
%
%
%
In our computation, $\tilde{\mbf{x}}_L$ can be obtained by integrating the ODE system numerically with proposed $\mbf{\theta}$ and initial values of $\mbf{x}_M$ and $\mbf{x}_L$.  Then, the target is to maximize the likelihood function $\rho(\mbf{x}_M,\mbf{\theta}|\mbf{y},\phi,\sigma,\gamma)$. 

The present algorithm is a combination of a Gaussian process based gradient matching and a least square optimization. In the GP gradient matching step, the Gaussian process model is first fitted by inferring the hyperparameter $\phi$. Secondly, the states and parameters are inferred using a one chain MCMC scheme on the density as in \cite{Wenk2019}. Finally, the parameters estimated above is set as initial guess in the least square optimization. The algorithm can be summarized as follows. 
\begin{table}[H]
\begin{tabular}{ll}  
\hline  
 &\textbf{Algorithm}\\
\hline
 & Input: $\mbf{y},\mbf{f}(\mbf{x},\mbf{\theta}),\gamma,N_{MCMC},N_{burnin},t,\sigma_s,\sigma_p$\\ 
& Step 1. Fit GP model to data\\
&  Step 2. Infer $\mbf{x}_M$, $\mbf{x}_L$ and $\mbf{\theta}$ using MCMC\\
& $S\leftarrow \emptyset$\\
& \emph{for} $i=1 \rightarrow N_{MCMC} + N_{burnin}$ \emph{do}\\
& $\hspace{2.0em}$\emph{for} each state \emph{do}\\
& $\hspace{4.0em}$Propose a new state value using a Gaussian distribution with standard \\
 & $\hspace{4.0em}$deviation $\sigma_s$\\
& $\hspace{4.0em}$Accept proposed value based on the density (Eq. \ref{joint-density-proof})\\
& $\hspace{4.0em}$Add current value to $\mathcal{T}$\\
& $\hspace{2.0em}$\emph{end for}\\
%
& $\hspace{2.0em}$\emph{for} each parameter \emph{do}\\
& $\hspace{4.0em}$ $\mathcal{T}\leftarrow\emptyset$\\
& $\hspace{4.0em}$Propose a new parameter value using a Gaussian distribution with standard\\
& $\hspace{4.0em}$deviation $\sigma_p$\\
& $\hspace{4.0em}$ Integrate $\mbf{x}_L$ with initial values and proposed parameters of ODEs.\\
& $\hspace{4.0em}$Accept proposed value based on the density (Eq. \ref{joint-density-proof})\\
& $\hspace{4.0em}$Add current value to $\mathcal{T}$\\
& $\hspace{2.0em}$\emph{end for}\\
& \emph{end for}\\
& Discard the first $N_{burnin}$ samples of $\mathcal{S}$\\
& Return $\mbf{x}_M,\mbf{x}_L,\mbf{\theta}$\\
& Step 3. Optimization using $\mbf{\theta}$ from Step 2 as initial guess.\\
\hline
\end{tabular}
\end{table}
In Step 1, the Gaussian process model is fitted to data by maximizing the $\log$ of marginal likelihood of the observations $\mbf{y}$ at times $\mbf{t}$
\begin{equation}
\log (\rho(\mbf{y}|\mbf{t},\phi,\sigma))=-\frac{1}{2}\mbf{y}^T(\mbf{C}_\phi+\sigma\mbf{I})^{-1}\mbf{y}-\frac{1}{2}\log|\mbf{C}_\phi+\sigma\mbf{I}|-\frac{n}{2}\log 2\pi ,    
\end{equation}
with respect to hyperparameters $\phi$ and $\sigma$. $\sigma$ is the standard deviation of the observation noise and $n$ is the amount of observations. 
The numerical integrals of $\tilde{\mbf{x}}_L$ in Step 2 are calculated only after each update of $\mbf{\theta}$. Step 3 is  to solve the following minimization problem,
\begin{equation}
\min_{\mbf{\theta}} \| \mbf{x}_M(\mbf{\theta})-\mbf{y}\|_{L^2(0,T)}^2.
\end{equation}
In the optimization process, gradient descent method is adopted where numerical gradient is used in each searching step. 
One advantage of doing optimization is its ability to obtain a more accurate result with less computational cost. In fact, with increased data the Gaussian noise can be balanced in the cost functional. But it requires proper initial guess of the parameters so as to avoid falling in local minima, whereas FGPGM has relatively less restrictions on the initial guess. However, for FGPGM a large amount of MCMC samplings are necessary to ensure the expectations of random variables make sense and it is hard to estimate the accuracy of the reconstructed solution. Therefore, if we combine these two methods, it is possible to use less MCMC sampling number to obtain a rough approximation of the parameters first and then adopt them as initial guess to obtain a least square optimization result.

\section{Experiments}

For the Gaussian regression step for observable variables, the code published alongside Wenk et al. (2019)\cite{Wenk2019} was used. The MCMC sampling part should then be adapted to the partial observation case according to the diagram provided above. In the following we refer FGPGM to the adapted FGPGM method for partial observation (Step 1 and Step 2 in the present algorithm).

\subsection{Lotka Volterra}
\begin{figure}[H] 
\begin{center}
\subfigure[$x_1$]
	{
	\includegraphics[width=0.31\textwidth]{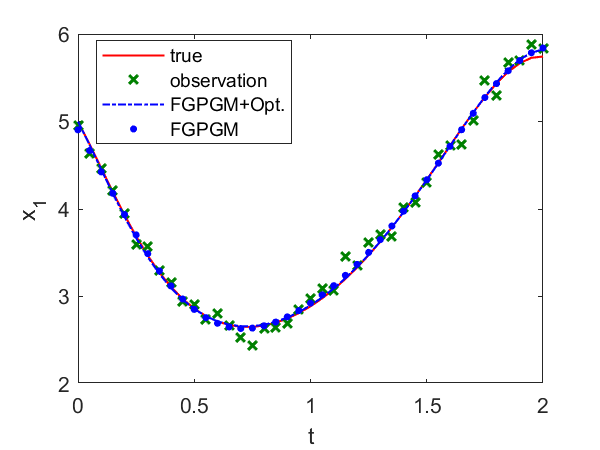}
	\label{eg1-1-1} 
	}
\subfigure[$x_2$]
	{
	\includegraphics[width=0.31\textwidth]{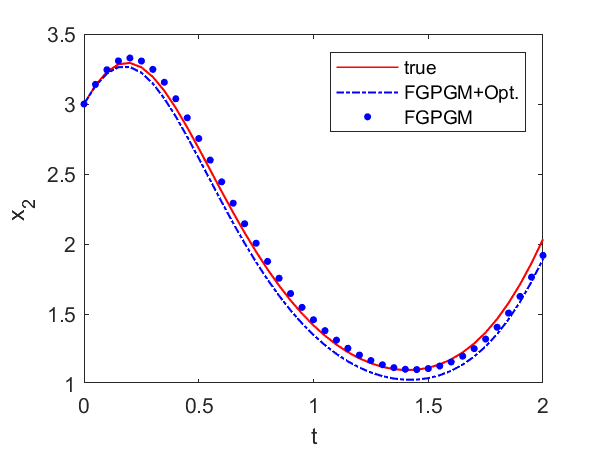}
	\label{eg1-1-2} 
	}
\subfigure[$\theta$]
	{
	\includegraphics[width=0.31\textwidth]{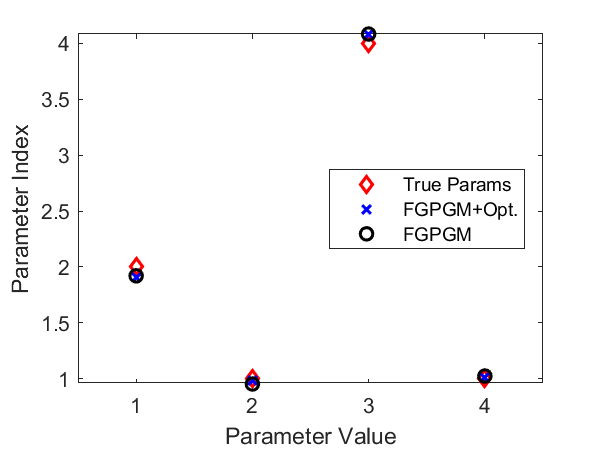}
	\label{eg1-1-theta} 
	}
\end{center}
  \caption{Reconstruction and inference results for the Lotka Volterra system, showing the state evolution over time and parameter distributions. $x_1$ is observable and $x_2$ is latent variable. The ground truth, FGPGM result, and result from combination of FGPGM and optimization are compared.}
\label{eg1-1} 
\end{figure}

\begin{figure}[H] 
\begin{center}
\subfigure[$x_1$]
	{
	\includegraphics[width=0.31\textwidth]{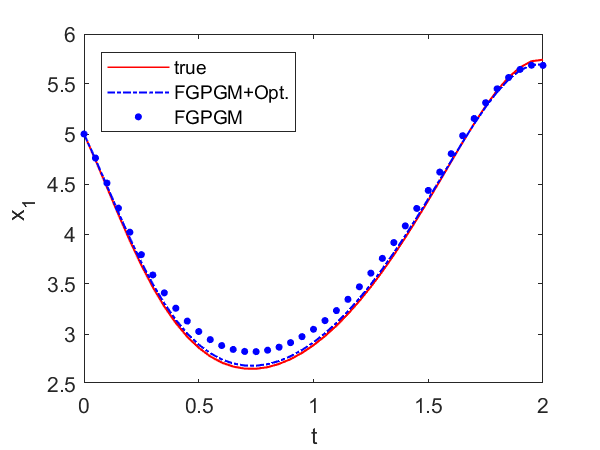}
	\label{eg1-2-1} 
	}
\subfigure[$x_2$]
	{
	\includegraphics[width=0.31\textwidth]{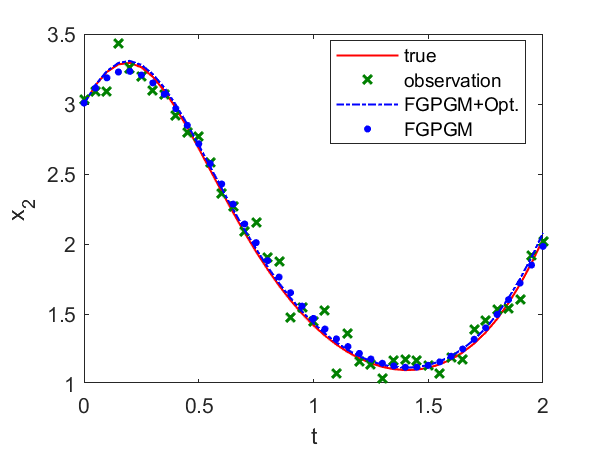}
	\label{eg1-2-2} 
	}
\subfigure[$\theta$]
	{
	\includegraphics[width=0.31\textwidth]{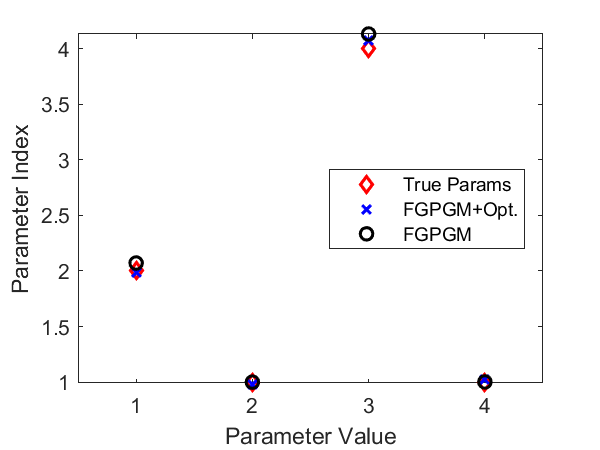}
	\label{eg1-2-theta} 
	}
\end{center}
  \caption{The state evolution over time for Lotka Volterra system and parameter inference results. $x_2$ is observable and $x_1$ is latent variable. The ground truth, FGPGM result, and result from combination of FGPGM and optimization are compared.}
\label{eg1-2} 
\end{figure}

The Lotka Volterra system was originally proposed in Lotka (1978)\cite{Lotka1978}. It was introduced to model the prey-predator interaction system whose dynamics are given by
\begin{align}
\dot{x}_1&= \theta_1 x_1(t)-\theta_2 x_1(t)x_2(t)\\
\dot{x}_2&=-\theta_3 x_2(t)+\theta_4 x_1(t)x_2(t),
\end{align}
where $\theta_1,\theta_2,\theta_3,\theta_4>0$. In the present work, the system was observed with one variable and the initial value of the other variable. The other setup is the same as Gorbach et al.(2017)\cite{Gorbach2017} and Wenk et al. (2019)\cite{Wenk2019}. The observed series are located in the time interval $[0,2]$ at 20 uniformly distributed observation times.  The initial values of the variables are $(5,3)$. The history of the observable variable is generated with numerical integration of the system with true parameters $\theta_1=2,\theta_2=1,\theta_3=4,\theta_4=1$, added by Gaussian noise with standard deviation 0.1. The RBF kernel was used for the Gaussian process. For the model noise we set $\gamma=3\times 10^{-1}$. The results with $x_1$ being observed is shown in Fig. \ref{eg1-1}. Those with observation of $x_2$ are given in Fig. \ref{eg1-2}. In the later case, we can see that the optimization process can improve the results from FGPGM, with the identified parameters being closer to the true values.

The sensitivities of the variables to the parameters are listed in Tab. \ref{error_table_1}. The sensitivity indexes at the true parameter set $\mbf{\theta_0}$ are defined as
\begin{equation}
S_{ij}=\frac{1}{\|x_i\|_{L^2(T_1,T_2)}}  \left\|\frac{\partial x_i(t;\mbf{\theta})}{\partial\theta_j}\right\|_{L^2(T_1,T_2)}(\mbf{\theta_0})
\label{sensitivity_index_def}
\end{equation}
which are normalized. It is approximated by numerical difference. It is shown that near the true parameter set, $\theta_1$ and $\theta_3$ are relatively less sensitive to the variables than other parameters. This explains that $\theta_1$ and $\theta_3$ are less accurate in the numerical test (see Fig. \ref{eg1-1-theta} and Fig. \ref{eg1-2-theta}).

\begin{table}[H] 
\begin{center} 
\begin{tabular}{ccc}  
\hline  
$S_{ij}$ & $x_1$ & $x_2$  \\  
\hline  
$\theta_1$  &  0.20  &  0.61  \\
$\theta_2$  &  0.52  &  1.13  \\
$\theta_3$  &  0.40  &  0.33  \\
$\theta_4$  &  1.27  &  0.98  \\
\hline  
\end{tabular} 
\end{center}
\caption{Sensitivity of each variable to parameters for Lotka Volterra system at $\mbf{\theta}=(2,1,4,1)$. The sensitivity index is defined Eq. \ref{sensitivity_index_def}.} 
\label{error_table_1} 
\end{table}

\begin{figure}[H] 
\begin{center}
\subfigure[$x_1$]
	{
	\includegraphics[width=0.31\textwidth]{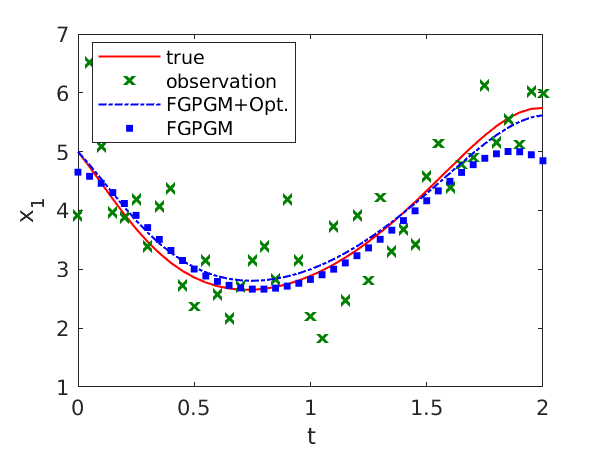}
	\label{eg1-1-1-noise} 
	}
\subfigure[$x_2$]
	{
	\includegraphics[width=0.31\textwidth]{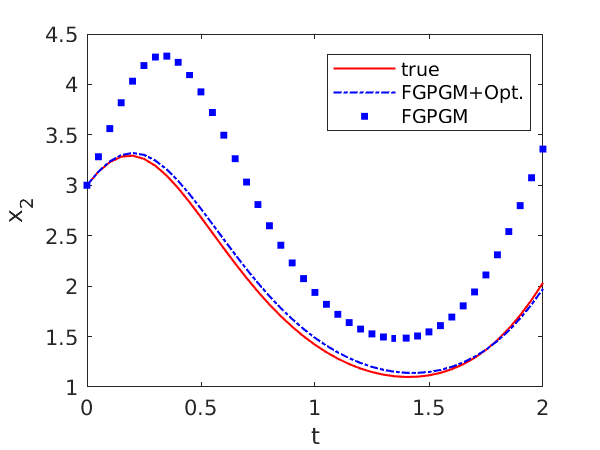}
	\label{eg1-1-2-noise} 
	}
\subfigure[$\theta$]
	{
	\includegraphics[width=0.31\textwidth]{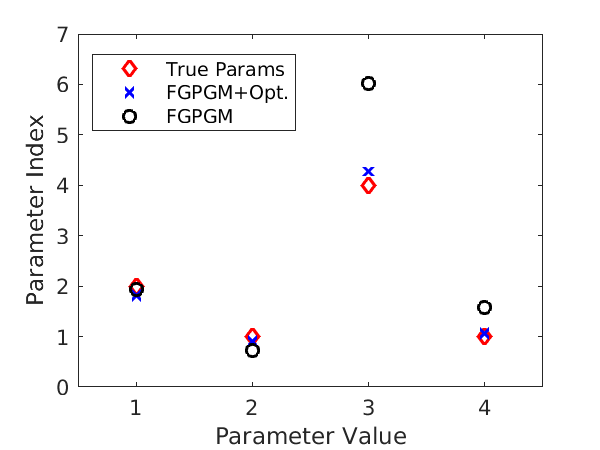}
	\label{eg1-1-theta-noise} 
	}
\end{center}
  \caption{Reconstruction and inference results for the Lotka Volterra system with $x_1$ being observable and $x_2$ latent. The ground truth, FGPGM result, and result from combination of FGPGM and optimization are compared. The observation noise has a standard deviation $std=0.5$ (large noise case).}
\label{eg1-1-noise} 
\end{figure}

\begin{figure}[H] 
\begin{center}
\subfigure[$x_1$]
	{
	\includegraphics[width=0.31\textwidth]{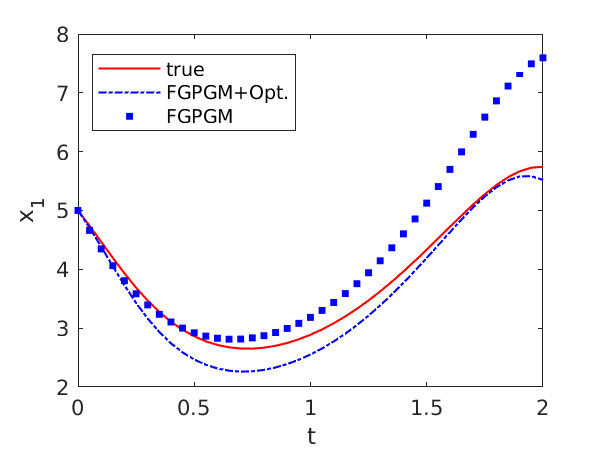}
	\label{eg1-2-1-noise} 
	}
\subfigure[$x_2$]
	{
	\includegraphics[width=0.31\textwidth]{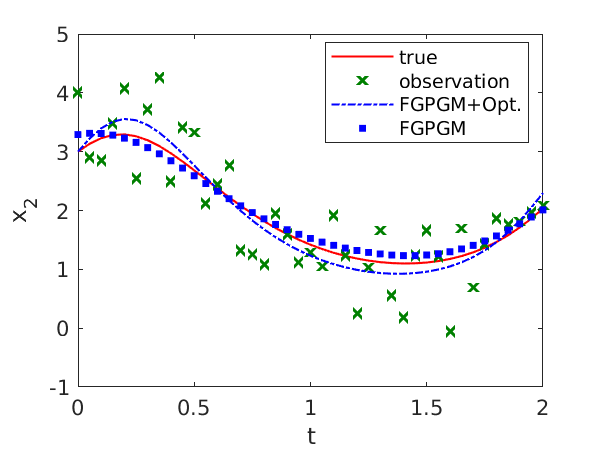}
	\label{eg1-2-2-noise} 
	}
\subfigure[$\theta$]
	{
	\includegraphics[width=0.31\textwidth]{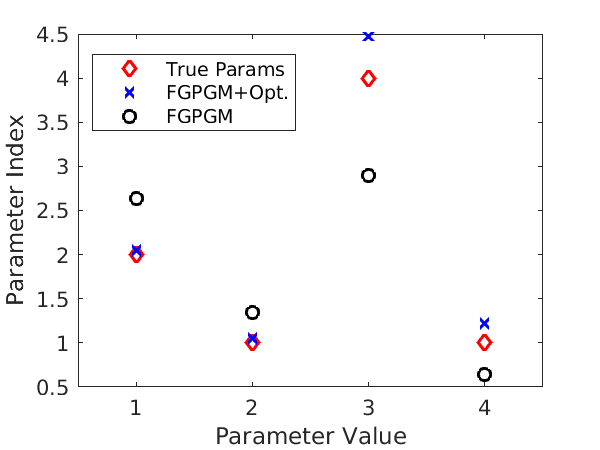}
	\label{eg1-2-theta-noise} 
	}
\end{center}
  \caption{The state evolution over time for Lotka Volterra system. $x_2$ is observable and $x_1$ is latent variable. The ground truth, FGPGM result, and result from combination of FGPGM and optimization are compared. The observation noise has a standard deviation $std=0.5$ (large noise case).}
\label{eg1-2-noise} 
\end{figure}

The cases with larger noise level (std=0.5) are shown in Fig. \ref{eg1-1-noise} and Fig. \ref{eg1-2-noise}, corresponding to $x_1$ and $x_2$ observations respectively. It can be seen that the prediction of the unknown variable can deviate far from the ground truth if we use FGPGM method only. The inferring of states and parameters can be improved after further applying the deterministic optimization.

\subsection{Spiky Dynamics}
This example is a system proposed by FitzHugh (1961) and Nagumo et al. (1962) for modeling the spike potentials in the giant squid neurons, which is abbreviated as FHN system. This system involves two ODEs with three parameters. The FHN system has notoriously fast changing dynamics due to its highly nonlinear terms. In the following numerical tests, the Matern 52 kernel was used and $\gamma$ was set to $3\times 10^{-1}$, the same as that in Wenk et al. (2019)\cite{Wenk2019}. We assume one of the two variables is observable, which was generated with $\theta_1=0.2$, $\theta_2=0.2$, $\theta_3=3$ and added by Gaussian noise with average signal-to-noise ratio $SNR=100$. There were 100 data points uniformly spaced in $[0,10]$. 
\begin{align}
&\dot{V}=\theta_1(V-\frac{V^3}{3}+R)\\
&\dot{R}=\frac{1}{\theta_1}(V-\theta_2+\theta_3 R)
\end{align}

\begin{table}[H] 
\begin{center} 
\begin{tabular}{ccc}  
\hline  
$S_{ij}$ & $x_1$ & $x_2$  \\  
\hline  
$\theta_1$  &  2.33  &  1.24  \\
$\theta_2$  &  0.44  &  0.31  \\
$\theta_3$  &  1.01  &  0.55  \\
\hline  
\end{tabular} 
\end{center} 
\caption{Sensitivity of each variable to parameters for FHN system at $\mbf{\theta}=(0.2,0.2,3.0)$. The sensitivity index is defined as Eq. \ref{sensitivity_index_def}.} 
\label{error_table_2} 
\end{table}

\begin{figure}[H] 
\begin{center}
\includegraphics[width=0.31\textwidth]{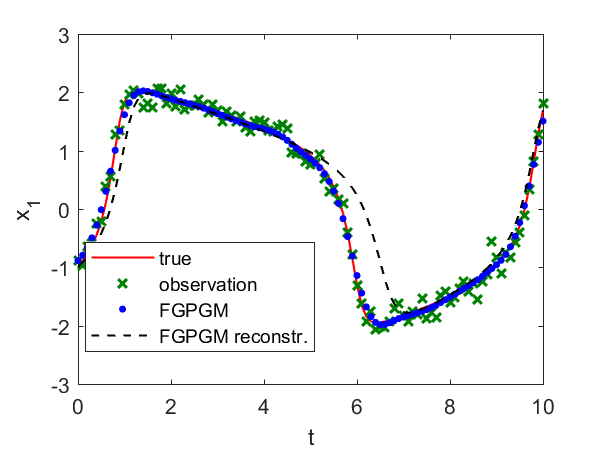}
\includegraphics[width=0.31\textwidth]{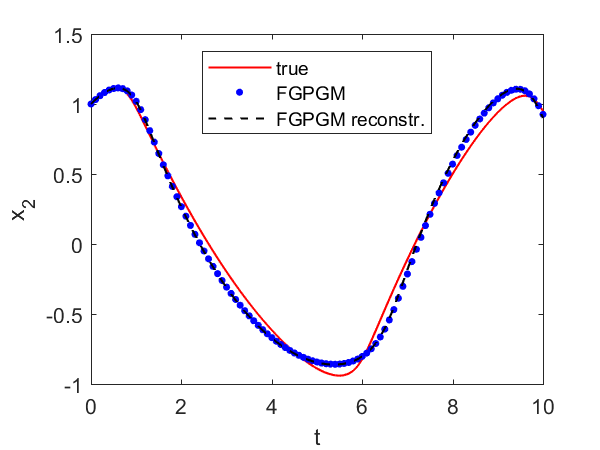}
\end{center}
\caption{Results for FHN system obtained from FGPGM method, without further optimization. $x_1$ is observable and $x_2$ is latent variable. The ground truth, FGPGM result, and reconstructed solution (integration of ODEs with inferred parameters) are compared.}
\label{eg3-rm2}
\end{figure}

\begin{figure}[H] 
\begin{center}
\subfigure[$x_1$]
	{
	\includegraphics[width=0.31\textwidth]{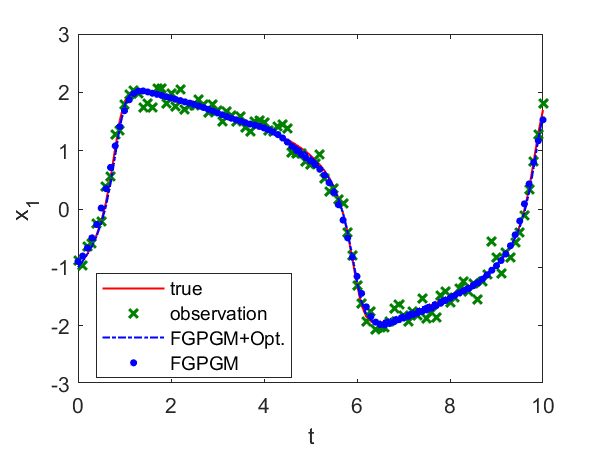}
	\label{eg3-1-1} 
	}
\subfigure[$x_2$]
	{
	\includegraphics[width=0.31\textwidth]{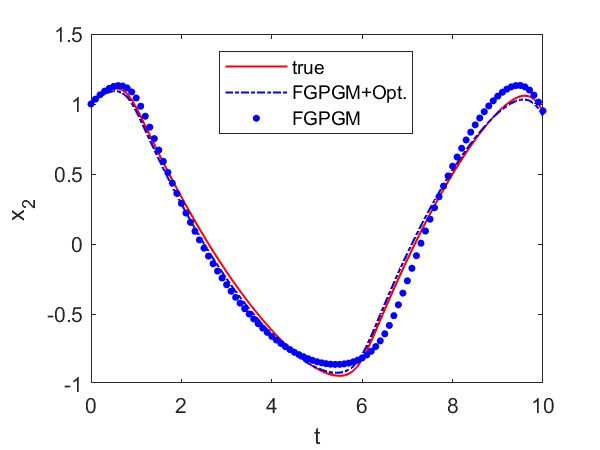}
	\label{eg3-1-2} 
	}
\subfigure[$\theta$]
	{
	\includegraphics[width=0.31\textwidth]{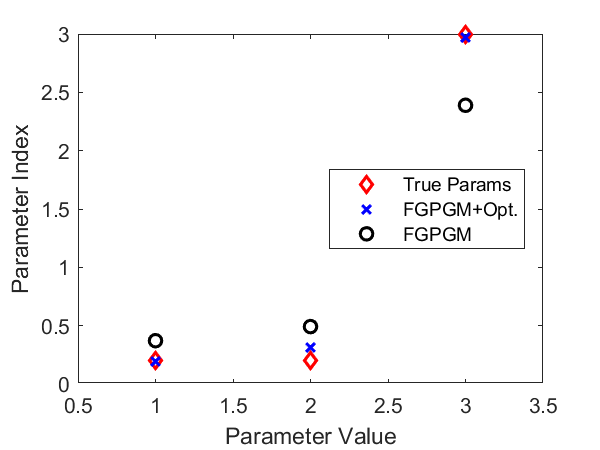}
	\label{eg3-1-theta} 
	}
\end{center}
  \caption{The state evolution over time and identified parameters for FHN system. $x_1$ is observable and $x_2$ is latent variable. The ground truth, FGPGM result, and result from combination of FGPGM and optimization are compared.}
\label{eg3-1} 
\end{figure}

In this case, if we merely use FGPGM step, the reconstructed solution corresponding to the identified parameters may deviate significantly from the true time series (see Fig. \ref{eg3-rm2}, where data of $x_1$ are observable). It was pointed out \cite{Wenk2019} that all GP based gradient matching algorithms lead to smoother trajectories than the ground truth. This becomes more severe with sparse observation. Thus a least square optimization after doing FGPGM may well reduce this effect (see Fig. \ref{eg3-1}). 

\begin{figure}[H] 
\begin{center}
\subfigure[$x_1$]
	{
	\includegraphics[width=0.31\textwidth]{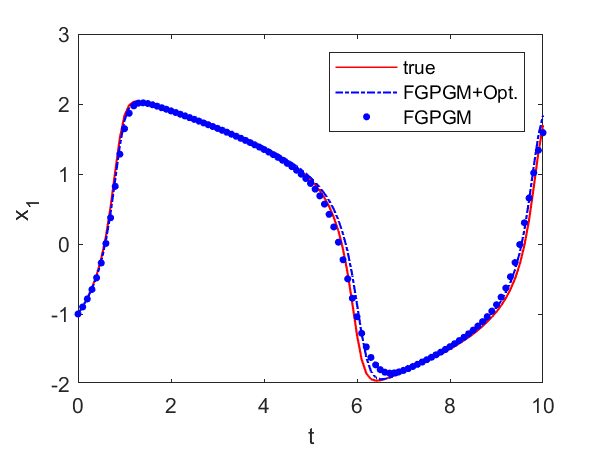}
	\label{eg3-2-1} 
	}
\subfigure[$x_2$]
	{
	\includegraphics[width=0.31\textwidth]{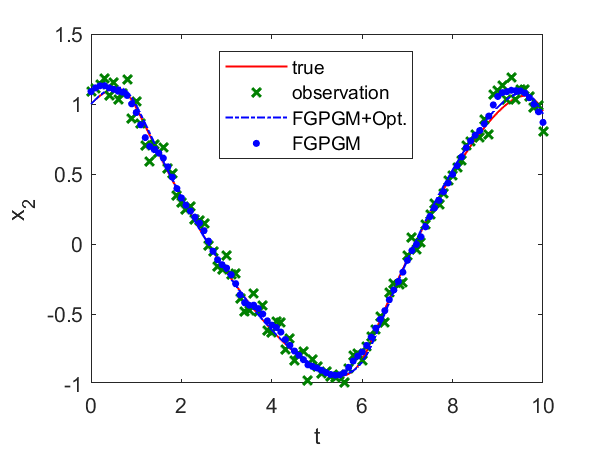}
	\label{eg3-2-2} 
	}
\subfigure[$\theta$]
	{
	\includegraphics[width=0.31\textwidth]{Eg_state_method1_params.png}
	\label{eg3-2-theta} 
	}
\end{center}
  \caption{The state evolution over time and identified parameters for FHN system. $x_2$ is observable and $x_1$ is latent variable. The ground truth, FGPGM result, and result from combination of FGPGM and optimization are compared.}
\label{eg3-2} 
\end{figure}

Fig. \ref{eg3-1} and Fig. \ref{eg3-2} present the results with single $x_1$ and $x_2$ observations respectively. In both cases the identified parameters are more accurate than using FGPGM only. From the sensitivity check in Tab. \ref{error_table_2}, it is expected that $\theta_1$ is most accurate because it is most sensitive among these three parameters, whereas $\theta_2$ is most insensitive and would be harder to be identified. The numerical results agree with that. It is worth mention that in the FGPGM step, only 3500 samplings were taken and the time for optimization step was much less than FGPGM step. This means the time needed for the whole process can be greatly saved compared with that in Wenk et al. 2019 \cite{Wenk2019}, where 100,000 MCMC samplings were implemented.  

\begin{figure}[H] 
\begin{center}
\includegraphics[width=0.31\textwidth]{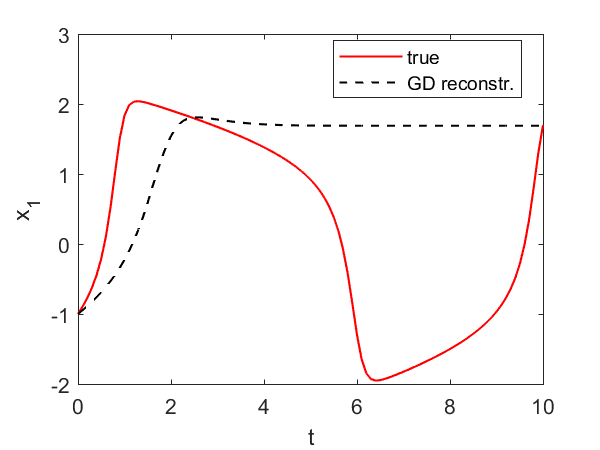}
\includegraphics[width=0.31\textwidth]{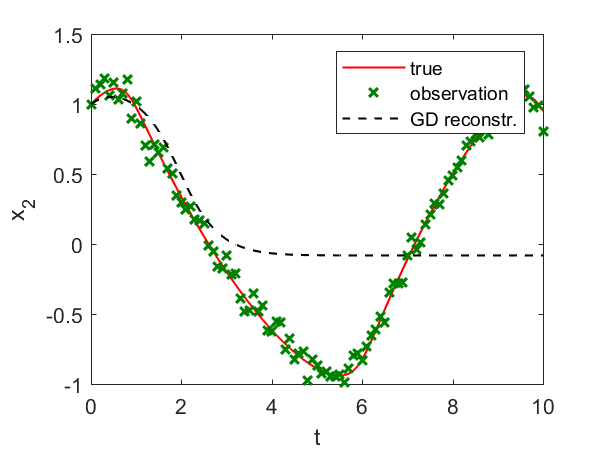}
\end{center}
\caption{Results of FHN system with $x_1$ being latent, obtained by merely using least square optimization with initial guess of parameters near a local minimum point.}
\label{eg3-rm1}
\end{figure}

In this example, we also notice that if we merely use least square optimization method, the local minimum effect would lead to reconstruction being far from the ground truth, which is even less robust than FGPGM method. For example, if we choose initial guess of the parameters near $(\theta_1,\theta_2,\theta_3)=(1.51,2.2,1.78)$ then the costfunctional will fall into the local minimum during gradient based search (see Fig. \ref{eg3-rm1}). The existence of many local minima in the full observation case has been pointed out in e.g., \cite{Esposito2000}\cite{Ramsay2007}. These results clearly illustrate the performance of the combination of FGPGM and least square optimization.

\subsection{Protein Transduction}
Finally the Protein Transduction system proposed in Vyshemisky and Girolami (2008) \cite{Vyshemirsky2008} was adopted to illustrate the performance of the method in ODEs with more equations. As mentioned in Dondelinger et al., 2013 \cite{Dondelinger2013}, it is notoriously difficult to fit with unidentifiable parameters. The system is described by 
\begin{align}
&\dot{S}=-\theta_1 S-\theta_2SR+\theta_3 R_S\\
&\dot{dS}=\theta_1 S\\
&\dot{R}=-\theta_2 SR +\theta_3 R_S +\theta_5\frac{R_{pp}}{\theta_6+R_{pp}}\\
&\dot{R}_S=\theta_2 SR-\theta_3 R_S-\theta_4 R_S\\
&\dot{R}_{pp}=\theta_4 R_S-\theta_5\frac{R_{pp}}{\theta_6+R_{pp}} .
\end{align}
The $\theta_6$ in this system is unidentifiable. We adopted the same experimental setup of Dondelinger et al. 2013 and Wenk et al. 2019 as follows. $\gamma=10^{-4}$ in FGPGM step. The observation were made at discrete times $[0,1,2,4,5,7,10,15,20,30,40,50,60,80,100]$. The initial condition was $[1,0,1,0,0]$ and the data were generated by numerical integrating the system under $\mbf{\theta}=[0.07,0.6,0.05,0.3,0.017,0.3]$, added by Gaussian noise with standard deviation 0.01.  A sigmoid kernel was used to deal with the logarithmically spaced observation times and the typically spiky form of the dynamics as in the previous papers.

\begin{table}[H] 
\begin{center} 
\begin{tabular}{cccccc}  
\hline  
$S_{ij}$ & $x_1$ & $x_2$ & $x_3$ & $x_4$ & $x_5$  \\  
\hline  
$\theta_1$  &  2.86  &  9.78  &  1.73 &  1.77  &  3.33 \\
$\theta_2$  &  0.70  &  0.98  &  0.22 &   0.59 &   0.41\\
$\theta_3$  &  1.35  &  2.11  &  0.47 &   0.92  &  0.90\\
$\theta_4$  &  0.26  &  0.43  &  0.03 &   2.64  &  0.62\\
$\theta_5$  &  1.53  &  2.58  & 24.48 &   0.90  & 49.38\\
$\theta_6$  &  0.04  &  0.07  &  0.60 &   0.02  &  1.21\\
\hline  
\end{tabular} 
\end{center}  
\caption{Sensitivity of each variable to parameters for Protein Transduction system at $\mbf{\theta}=[0.07,0.6,0.05,0.3,0.017,0.3]$. The sensitivity index is defined as Eq. \ref{sensitivity_index_def}.} 
\label{error_table_3} 
\end{table}

\begin{figure}[H] 
\begin{center}
\subfigure[$x_1$]
	{
	\includegraphics[width=0.31\textwidth]{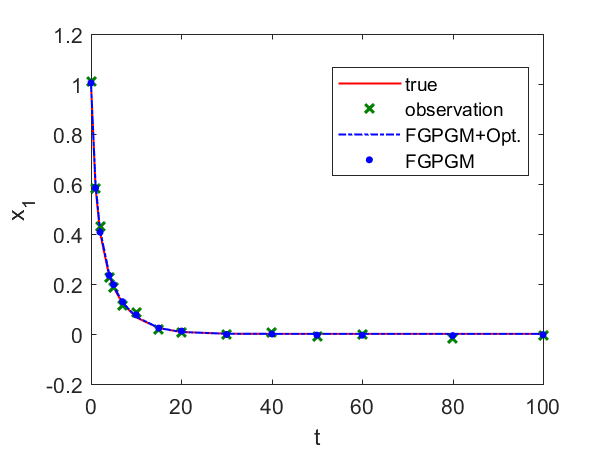}
	\label{eg3-lack3-1} 
	}
\subfigure[$x_2$]
	{
	\includegraphics[width=0.31\textwidth]{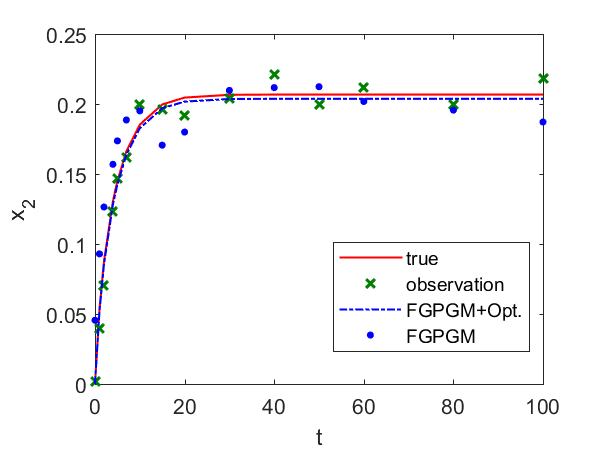}
	\label{eg3-lack3-2} 
	}
\subfigure[$x_3$]
	{
	\includegraphics[width=0.31\textwidth]{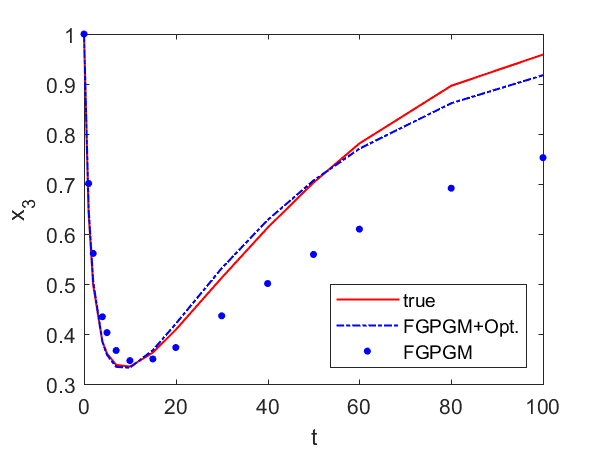}
	\label{eg3-lack3-3} 
	}
\subfigure[$x_4$]
	{
	\includegraphics[width=0.31\textwidth]{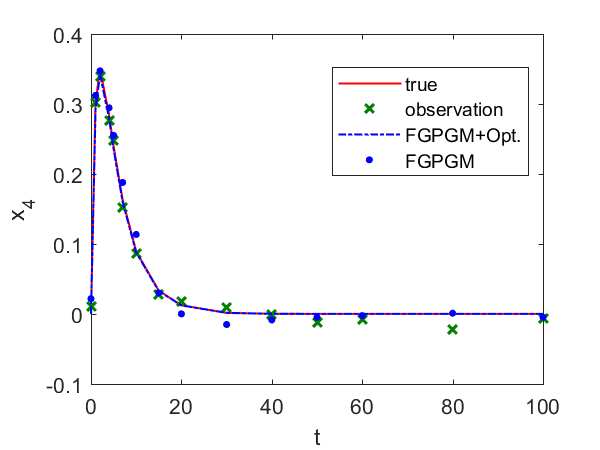}
	\label{eg3-lack3-4} 
	}
\subfigure[$x_5$]
	{
	\includegraphics[width=0.31\textwidth]{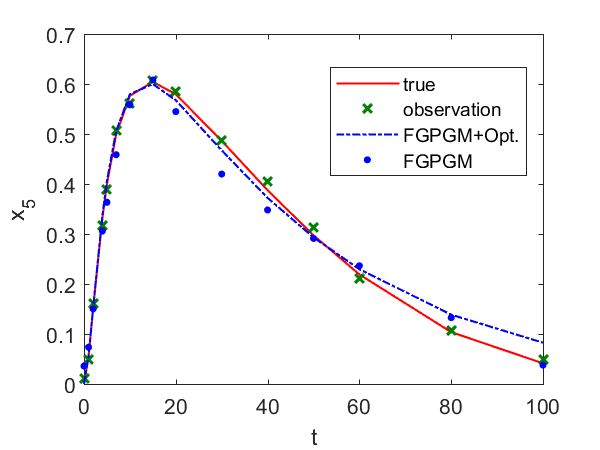}
	\label{eg3-lack3-5} 
	}
\subfigure[$\theta$]
	{
	\includegraphics[width=0.31\textwidth]{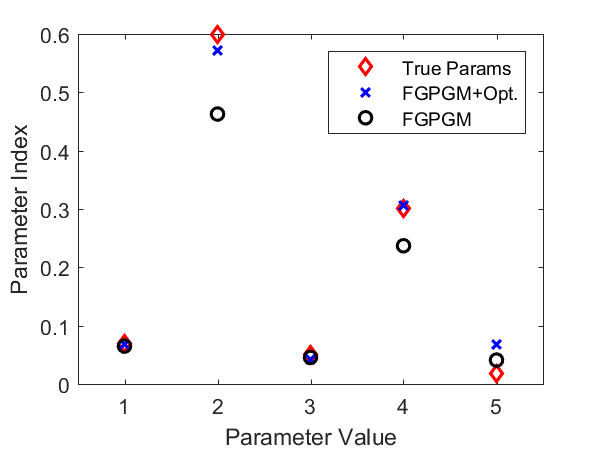}
	\label{eg3-lack3-6} 
	}
\end{center}
  \caption{The state evolution over time and inferred parameters for protein transduction system. $x_3$ is unknown and other variables are observable. The ground truth, FGPGM result, and result from combination of FGPGM and optimization are compared.}
\label{eg3_lack3} 
\end{figure}

Fig. \ref{eg3_lack3} gives the result with $x_3$ ($R$) being unobserved. In fact, the situations with one of other variables being unknown have better results than the case illustrated here, which will not be presented here. We can see that $x_3$ was not well fitted by merely using FGPGM step, whereas the combination of FGPGM and optimization generated satisfactory result, with the parameters $\theta_2$ and $\theta_4$ being significantly improved. The sensitivity check is summarized in Tab. \ref{error_table_3}, from which we can see that $\theta_2$ is less sensitive and thereby harder to infer accurately. The error of $\theta_5$ may be affected by the value of the unidentifiable parameter $\theta_6$.

\begin{figure}[H] 
\begin{center}
\subfigure[$x_1$]
	{
	\includegraphics[width=0.31\textwidth]{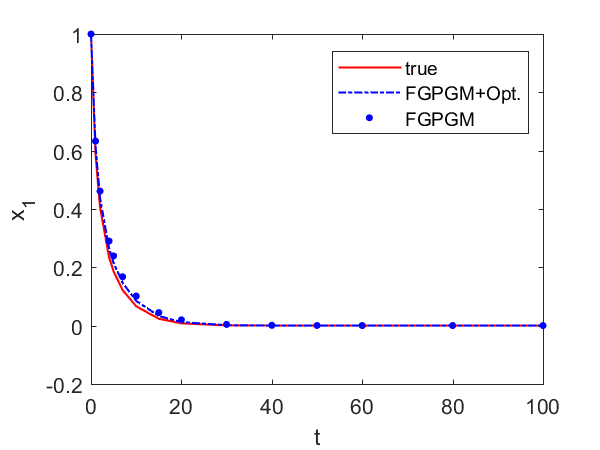}
	\label{eg2-lack13-1} 
	}
\subfigure[$x_2$]
	{
	\includegraphics[width=0.31\textwidth]{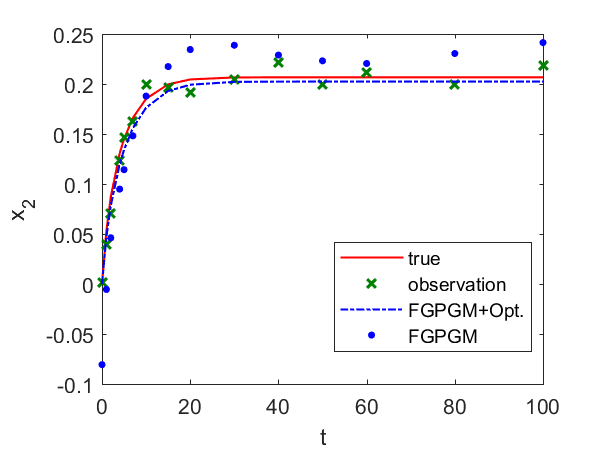}
	\label{eg2-lack13-2} 
	}
\subfigure[$x_3$]
	{
	\includegraphics[width=0.31\textwidth]{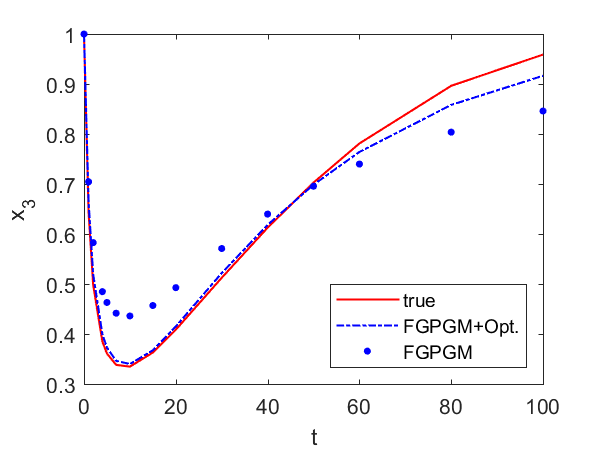}
	\label{eg2-lack13-3} 
	}
\subfigure[$x_4$]
	{
	\includegraphics[width=0.31\textwidth]{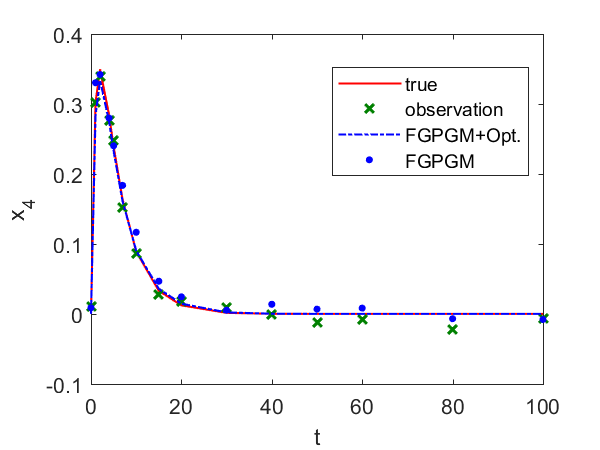}
	\label{eg2-lack13-4} 
	}
\subfigure[$x_5$]
	{
	\includegraphics[width=0.31\textwidth]{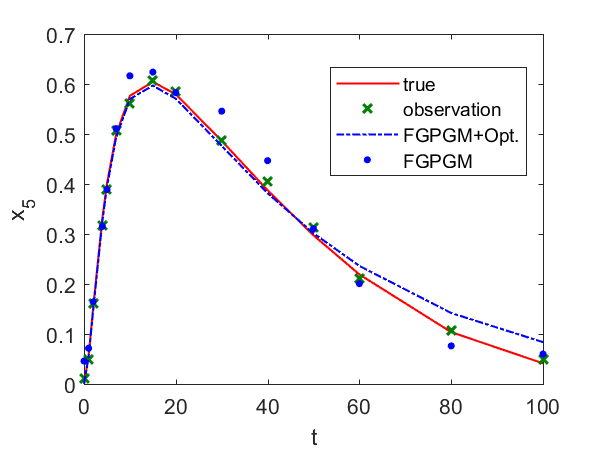}
	\label{eg2-lack13-5} 
	}
\subfigure[$\theta$]
	{
	\includegraphics[width=0.31\textwidth]{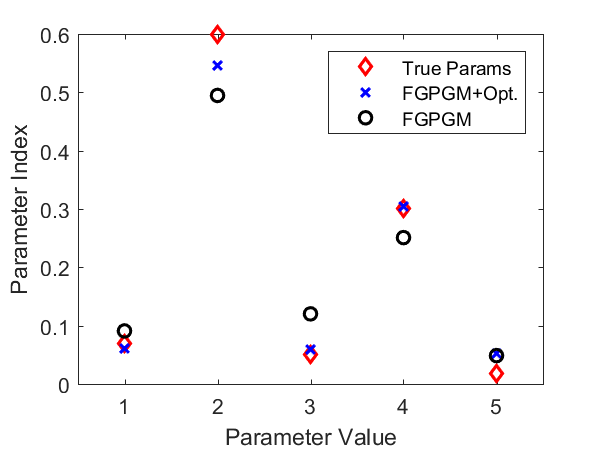}
	\label{eg2-lack13-6} 
	}
\end{center}
  \caption{The state evolution over time and inferred parameters for protein transduction system. $x_1$ and $x_3$ are unknown and other variables are observable. The ground truth, FGPGM result, and result from combination of FGPGM and optimization are compared.}
\label{eg2_lack13} 
\end{figure}

It would also be of interest to see the performance of the method for the cases with more latent variables. In this model, although $dS$ is not involved in equations for other variables, the data of $dS$ helps infer $\theta_1$. We also notice that $\dot{R}+\dot{R_S}+\dot{R}_{pp}=0$. If $S$ and $dS$ are both missing, it is impossible to identify $\theta_1$. Therefore, in the following test we choose data of $dS$, $R_s$ and $R_{pp}$ as observations. The data has a Gaussian noise with standard deviation 0.01, the same as the previous case with one latent variable. It can be seen from Fig. \ref{eg2_lack13} that the result from FGPGM step is worse than the case with only one latent variable, but the final reconstruction of latent variables and parameter identification is not significantly different from the case with one latent variable. 

\section{Discussion}
In the work, we proposed an effective method for parameter inference of coupled ODE systems with partially observable data. Our method is based on previous work known as FGPGM \cite{Wenk2019}, which avoids product of experts heuristics. In order to  improve the accuracy and efficiency of the method, we also use Least Square optimization in our computation. 
 Due to the existence of latent variables, numerical integration is necessary for computation of the likelihood function in FGPGM method, which increases computational cost.  The Least Square optimization allows us to greatly reduce the sampling number. In our numerical tests, we show that  the  sampling number in the FGPGM step is only 10\% of that suggested in literature. It is worth noting that  conventional least square optimization method requires good initial guess, which is not the case in our approach. Our numerical examples illustrated here demonstrate the feasibility of the proposed method for parameter inference with partial observations. 
\section*{Acknowledgments}
This work was supported in part by National Science Foundation of China  (NSFC: No. 11971121), Nature Science and Research Council (NSERC) of Canada and the Fields Institute for Research in Mathematical Sciences. The authors also express the gratitude to Prof. Xin Gao, Nathan Gold and other members of the Fileds CQAM Lab on Health Analytics and Modelling for very beneficial discussions.

\addcontentsline{toc}{section}{

\appendix
\section{Preliminaries}\label{appendix-preliminaries}
In the following we list some preliminaries on derivatives of a Gaussian process that are used in this work, the proofs can be find in, e.g., \cite{Papoulis}\cite{Rasmussen}\cite{Wenk2019}. Denote a random process $X_t$, its realization $x$ and its time derivative $\dot{X}_t$.
\begin{definition}(\cite{Papoulis})
	The random variable $X_n$ converges to $x$ in the first-mean sense (limit in mean) if for some $X$,
\begin{equation}
\lim_{n\rightarrow \infty}\mathbb{E}(|X_n-X|)=0.
\end{equation}
\end{definition}
 
\begin{definition}
	The stochastic process $X_t$ is first-mean differentiable if for some $\dot X_t$
	\begin{equation}
	\lim_{\delta t \rightarrow 0 }\mathbb{E} \left|\frac{X_{t+\delta t}-X_t}{\delta t} - \dot X_t\right| = 0. 
	\end{equation}
\end{definition}

\begin{definition}
	For given random variable $X$, the moment generating function (MGF) is defined by 
	\begin{equation}
	\Phi_X(t)= E[\exp(Xt)]=\int_{-\infty}^{\infty} \exp(xt)\rho(x)dx.
	\end{equation}
\end{definition}
\begin{proposition}
	If $\Phi_X(t)$ is the MGF, then
	\begin{enumerate}
		\item $\frac{d\Phi_X}{dt}|_{t=0} = m_i,$ 
		where $m_i$ is the $i_{th}$ moment of $X$. 
		\item Let  $X$ and  $Y$ be two random variables.   X and  Y have the same distribution if and only if they have the same MGFs.
		\item we say $X\sim N(\mu,\sigma^2)$ if and only if  $\Phi_X(t) = \exp^{\frac{\sigma^2t^2}{2}+\mu t}$.
		\item If $X$ and $Y$ are two random variable, then the MGF $\Phi_{X+Y}(t)=\Phi_X(t)\Phi_Y(t).$  
	\end{enumerate}
\end{proposition}
By the above propositions, one has
\begin{lemma}
	 If $X,Y$ are two independent Gaussian random variables with means $\mu_X, \mu_Y$ and covariances $\sigma_X^2, \sigma_Y^2$, then  $X+Y$ is a Gaussian random variable with mean $\mu_X+\mu_y$ and covariance $\sigma_X^2+\sigma_y^2$. 
\end{lemma}

 \begin{definition}(\cite{Rasmussen})
A real-valued stochastic process $\{X_t\}_{t\in T}$, where $T$ is an index set, is a Gaussian process if all the finite-dimensional distributions are a multivariate normal distribution. That is, for any choice of distinct values $t_1,t_2, \dots t_N\in T$, the random vector $\bX = (X_{t_1},\dots,X_{t_N})^T$ has a multivariate normal distribution with joint Gaussian probability density function given by 
\begin{equation}
\rho_{X_{t_1}X_{t_2}\dots X_{t_N}} (x_{t_1},\dots,x_{t_N}) = \frac{1}{(2\pi)^{N/2}det(\bSigma)^{1/2}}\exp\left(-\frac 1 2 (\bx-\bmu_{\bX})^T\bSigma^{-1}(\bx-\bmu_{\bX})\right).
\end{equation}
where the mean vector is defined as
 \begin{eqnarray}
 (\bmu_{\bX})_i = \mathbb{E}[\bX_{t_i}]
 \end{eqnarray}  and covariance matrix $(\bSigma)_{ij} = cov (X_{t_i},X_{t_j})$.
\end{definition}

	The Gaussian processes only depend on the mean and covariance functions. Usual covariance functions could be Squared exponential $cov(X_{t_i},X_{t_j})=k_{\phi}(t_i,t_j) = \exp (-\frac 1 {2l^2} |t_i-t_j|^2)$, where $l$ is a hyperparameter and represents the nonlocal interaction length scale.

Let $t_0, \delta t\in R$, and $X_t$ be a Gaussian Process with constant mean $\mu$ and kernel function $k_{\phi}(t_1,t_2)$, assumed to be first-mean differentiable. Then $X_{t_0+\delta t}$ and $X_{t_0}$ are jointed Gaussian distributed
\begin{equation}
\left[\begin{array}{c}
X_{t_0}\\
X_{t_0+\delta t}
\end{array}
\right]\sim \mathcal{N}\left(
\left[\begin{array}{c}
\mu\\
\mu
\end{array}
\right], \bSigma
\right)
\end{equation}
with density function 
\begin{equation}
\rho(x_{t_0},x_{t_0+\delta t})= \frac{1}{2\pi\det(\bSigma)^{1/2}}\exp\left({-\frac 1 2\left[\begin{array}{c}
	x_{t_0}-\mu\\
	x_{t_0+\delta t}-\mu  
	\end{array}
	\right]^T \bSigma^{-1}
\left[\begin{array}{c}
x_{t_0}-\mu\\
x_{t_0+\delta t} -\mu 
\end{array}
\right]
 }\right)
\end{equation}
where 
\begin{equation}
\bSigma = \left(\begin{array}{cc}
k_{\phi}(t_0,t_0) &k_{\phi}(t_0,t_0+\delta t)\\
k_{\phi}(t_0+\delta t,t_0)&k_{\phi}(t_0+\delta t,t_0+\delta t)
\end{array}
\right).
\end{equation}
If we define linear transformation 
\begin{eqnarray}
\mathbf{T}= \left(\begin{array}{cc}
1&0\\
 -\frac 1 {\delta t} &\frac 1 { \delta t}
\end{array}
\right),
\end{eqnarray}
then we have 
\begin{equation}
\left[\begin{array}{c}
X_{t_0}\\
\frac{X_{t_0+\delta t}-X_{t_0}}{\delta t} 
\end{array}
\right] =\mathbf{T}\left[\begin{array}{c}
X_{t_0}\\
 X_{t_0+\delta t} 
\end{array}
\right] \sim \mathcal{N}\left(
\left[\begin{array}{c}
\mu\\
0
\end{array}
\right],\mathbf{T} \bSigma\mathbf{T}^T
\right)
\end{equation}
i.e. 
\begin{equation}
\rho(X_{t_0}, \frac{X_{t_0+\delta t}-X_{t_0}}{\delta t}) = \mathcal{N}\left(
\left[\begin{array}{c}
\mu\\
0
\end{array}
\right],\mathbf{T} \bSigma\mathbf{T}^T\right)
\end{equation}
where 
\begin{eqnarray}
\mathbf{T} \bSigma\mathbf{T}^T= \left(
\begin{array}{cc}
k_{\phi}(t_0,t_0)&\frac{k_{\phi}(t_0,t_0+\delta t)-k_{\phi}(t_0,t_0)}{\delta t_0}\\
\frac{k_{\phi}(t_0+\delta t_0,t_0)-k_{\phi}(t_0,t_0)}{\delta t}&\frac{\frac{k_{\phi}(t_0+\delta t_0,t_0+\delta t)-k_{\phi}(t_0,t_0+\delta t)}{\delta t_0}-\frac{k_{\phi}(t_0+\delta t,t_0)-k_{\phi}(t_0,t_0)}{\delta t_0}}{\delta t}
\end{array}
\right).
\end{eqnarray}

The above derivation shows that $X_{t_0}$ and $\frac{X_{t_0+\delta t}-X_{t_0}}{\delta t} $
are jointly Gaussian distributed. Using the definition of $first-mean$ differential and the fact that $rth-mean$ convergence implies convergence in distribution, it is clear that $X_{t_0}$ and $\dot X_{t_0}$  are jointly Gaussian
\begin{equation}
\left[\begin{array}{c}
X_{t_0}\\
\dot X_{t_0}
\end{array}
\right] \sim \mathcal{N}\left(
\left[\begin{array}{c}
\mu\\
0
\end{array}
\right], \left[\begin{array}{cc}
 k_{\phi}(t_0,t_0)&\frac{\partial k_{\phi}(a,b)}{\partial b}|_{a=t_0,b=t_0}\\
\frac{\partial k_{\phi}(a,b)}{\partial a}|_{a=t_0,b=t_0}& \frac{\partial^2 k_{\phi}(a,b)}{\partial a\partial b}|_{a=t_0,b=t_0}
\end{array}\right]
\right).
\end{equation} 

In general, $ \bX = (X_{t_1},\dots,X_{t_k})^T$  and $\dot{\bX}= (\dot X_{t_1},\dots,\dot X_{t_k})^T$ are jointly Gaussian 
\begin{equation}
\left[\begin{array}{c}
\bX\\
\dot\bX
\end{array}
\right]   \sim \mathcal{N}\left(
\left[\begin{array}{c}
\bmu\\
\mbf{0}
\end{array}
\right], 
  \left[\begin{array}{cc}
\bC_{\phi}(\bX,\bX) & \bC_{\phi}(\bX,\dot{\bX})\\
\bC_{\phi}(\dot{\bX},\bX) &\bC_{\phi}(\dot{\bX},\dot{\bX})
\end{array}
\right] 
\right).
\end{equation}
Here $(C_{\phi}(\mathbf{a},\mathbf{b}))_{ij} = cov(a_i,b_j)$ is the covariance between between $a_i$ and $b_j$, and  predefined kernel matrix of Gaussian process.
By   linearity of the covariance operator and the predefined kernel function $k_{\phi}(a,b)$, we have 
\begin{eqnarray}
C_{\phi}(X_{t_i},\dot X_{t_j})  = \frac{\partial k_{\phi}(a,b)}{\partial b}|_{a=t_i,b=t_j},\\
C_{\phi}(\dot X_{t_i}, X_{t_j}) = \frac{\partial k_{\phi}(a,b)}{\partial a}|_{a=t_i,b=t_j}, \\
C_{\phi}(\dot X_{t_i}, \dot X_{t_j}) =\frac{\partial^2 k_{\phi}(a,b)}{\partial a\partial b }|_{a=t_i,b=t_j}.
\end{eqnarray}
\begin{lemma}(Matrix Inversions Lemma)
	Let $\bSigma$ be a $p\times p-matrix$ ($p=n+m$):
	\begin{equation}
	\bSigma = \left[\begin{array}{cc}
	\bSigma_{11}& \bSigma_{12}\\
	\bSigma_{21}&\bSigma_{22}
	\end{array}\right]
	\end{equation} 
where the sum-matrices have dimension $n\times n$, $n\times m$, etc. Suppose $\bSigma, \bSigma_{11}, \bSigma_{22}$ are non-singular; and  partition the inverse in the same way as $\bSigma$,
  	\begin{equation}
  \bLambda =\bSigma^{-1}= \left[\begin{array}{cc}
  \bLambda_{11}& \bLambda_{12}\\
  \bLambda_{21}&\bLambda_{22}
  \end{array}\right].
  \end{equation} 
Then \begin{eqnarray}\left\{\begin{array}{l}
\bLambda_{11} = (\bSigma_{11}-\bSigma_{12}\bSigma_{22}^{-1}\bSigma_{21})^{-1}\\
\bLambda_{12} = -(\bSigma_{11}-\bSigma_{12}\bSigma_{22}^{-1}\bSigma_{21})^{-1}\bSigma_{12}\bSigma_{22}^{-1}\\
\bLambda_{21}=-(\bSigma_{22}-\bSigma_{21}\bSigma_{11}^{-1}\bSigma_{12})^{-1}\bSigma_{21}\bSigma_{11}^{-1}\\
\bLambda_{22} = (\bSigma_{22}-\bSigma_{21}\bSigma_{11}^{-1}\bSigma_{12})^{-1}.\end{array}\right.
\end{eqnarray}
\end{lemma}

\begin{lemma}\label{conditionalmuandvariance}
(Conditional Gaussian distributions) Let   $\bX\in\mathbb{R}^D$, $\bY\in\mathbb{R}^M$,    be  jointly Gaussian random vectors with distribution  
\begin{equation}
\left[\begin{array}{c}
\bX\\
\bY
\end{array}
\right]\sim \mathcal{N}\left(
\bmu, \bSigma
\right)
\end{equation}

 where 
\begin{equation}
\bmu = \left[\begin{array}{c}
\bmu_X\\
\bmu_Y
\end{array}\right],
\bSigma =  \left[\begin{array}{cc}
\bSigma_{XX}&\bSigma_{XY}\\
\bSigma_{YX}& \bSigma_{YY}
\end{array}\right].
\end{equation}
Then the conditional Gaussian distributions density functions are 
\begin{eqnarray}
\rho_{Y|X}(\by|\bx) =\frac{\rho_{XY}(\bx,\by)}{\rho_{X}(\bx)}= \frac{1}{(2\pi)^{\frac{M+D}{2}}\det(\Sigma_{Y|X})^{1/2}}\exp{(\by-\bmu_{Y|X})^T\bSigma_{Y|X}^{-1}(\by-\bmu_{Y|X})}
\end{eqnarray}

where 
\begin{eqnarray}
\bmu_{Y|X}=\bmu_Y+\bSigma_{YX}\bSigma_{XX}^{-1}(\bx-\bmu_X),\\ \bSigma_{Y|X}=\bSigma_{YY}-\bSigma_{YX}\bSigma_{XX}^{-1}\bSigma_{XY}.
\end{eqnarray}
\end{lemma}

According to above Lemma,  we have the condition distribution  

\begin{lemma}
\begin{equation}
\rho(\dot{\mbf{x}}|\mbf{x}) \sim \mathcal{N}(\mbf{D}(\bx-\bmu_X), \mbf{A})
\end{equation} 
where
\begin{align}
   &\mbf{D}= \bC_{\phi}(\dot\bX,\bX)\bC_{\phi}(\bX,\bX)^{-1}\\
   &\mbf{A}=\bC_{\phi}(\dot\bX,\dot\bX)-\bC_{\phi}(\dot\bX,\bX)\bC_{\phi}(\bX,\bX)^{-1}\bC_{\phi}(\bX,\dot\bX)
\end{align}
\label{conditional-Gaussian-distribution-lemma}
\end{lemma}

\section{Proof of Theorem \ref{jointdensity-partial}}\label{appendix-proof}
\begin{proof} The joint density over all variables in Fig.\ref{Numerical-harmonic-measure} can be represented as

\begin{align}
 &\rho(\mbf{x}_M,\dot{\mbf{x}}_M,\mbf{y},\mbf{x}_L,F_M,\bar{F}_M,\mbf{\theta}|\phi,\sigma,\gamma)\notag\\
=&  \rho_{GP}(\mbf{x}_M,\dot{\mbf{x}}_M,\mbf{y}|\phi,\sigma)\rho_{ODE}(F_M,\bar{F}_M,\theta,\mbf{x}_L|\mbf{x}_M,\dot{\mbf{x}}_M,\gamma)
\end{align}
\begin{equation}
    \rho_{GP}(\mbf{x}_M,\dot{\mbf{x}}_M,\mbf{y}|\phi,\sigma)=\rho(\mbf{x}_M,\phi)\rho(\mbf{y}|\mbf{x}_M,\sigma)\rho(\dot{\mbf{x}}_M|\mbf{x}_M,\phi)
\end{equation}

\begin{align}
    &\rho_{ODE}(F_M,\bar{F}_M,\theta,\mbf{x}_L|\mbf{x}_M,\dot{\mbf{x}}_M,\gamma)\notag\\
    =&\rho(\mbf{\theta})\rho(F_M,\bar{F}_M,\mbf{x}_L|\mbf{\theta},\mbf{x}_M,\dot{\mbf{x}}_M,\gamma)\nonumber\\
    =&\rho(\mbf{\theta})\rho(\mbf{x}_L|\mbf{\theta},\mbf{x}_M)\rho(F_M,\bar{F}_M|  \mbf{x}_L,\mbf{\theta},\mbf{x}_M,\dot{\mbf{x}}_M,\gamma)\nonumber\\
    =&\rho(\mbf{\theta}) \delta( \tilde{\mbf{x}}_L(\mbf{\theta},\mbf{x}_M)-\mbf{x}_L) \rho(F_M,\bar{F}_M|  \mbf{x}_L,\mbf{\theta},\mbf{x}_M,\dot{\mbf{x}}_M,\gamma)\nonumber\\
    =&\rho(\mbf{\theta}) \rho (F_M|\mbf{\theta},\mbf{x}_M,\tilde{\mbf{x}}_L(\mbf{x}_M,\mbf{\theta}))\rho(\bar{F}_M|\dot{\mbf{x}}_M,\gamma)\delta(F_M-\bar{F}_M)\nonumber\\
    =&\rho(\mbf{\theta})\delta(\mbf{f}_M(\mbf{\theta},\mbf{x}_M,\tilde{\mbf{x}}_L(\mbf{x}_M,\mbf{\theta}))-F_M)\mathcal{N}(F_M|\dot{\mbf{x}}_M,\gamma\mbf{I})\nonumber\\
    =&\rho(\mbf{\theta})\mathcal{N}(\mbf{f}_M(\mbf{\theta},\mbf{x}_M,\tilde{\mbf{x}}_L(\mbf{x}_M,\mbf{\theta}))|\dot{\mbf{x}}_M,\gamma\mbf{I}),
\end{align}
by which $\rho_{ODE}$ is independent of $F_M,\bar{F}_M,\mbf{x}_L$. $\tilde{\mbf{x}}_L$ is deterministically decided by $\mbf{x}_M,\mbf{\theta}$ through integration. Using Lemma \ref{conditional-Gaussian-distribution-lemma}, we have
\begin{align}
    &\rho(\mbf{x}_M,\mbf{\theta},\mbf{y}|\phi,\sigma,\gamma)=\notag\\ &\rho(\mbf{\theta})\mathcal{N}(\mbf{x}_M|\mbf{0},\mbf{C}_\phi)\mathcal{N}(\mbf{y}|\mbf{x}_M,\sigma^2\mbf{I})\mathcal{N}(\dot{\mbf{x}}_M|\mbf{D}\mbf{x}_M,\mbf{A}) \mathcal{N}(\mbf{f}_M(\mbf{\theta},\mbf{x}_M,\tilde{\mbf{x}}_L(\mbf{x}_M,\mbf{\theta}))|\dot{\mbf{x}}_M,\gamma\mbf{I}).
\end{align}
Integrating $\dot{\mbf{x}}_M$ out yields
\begin{align}
    &\rho(\mbf{x}_M,\mbf{\theta},\mbf{y}|\phi,\sigma,\gamma)=\notag\\ &\rho(\mbf{\theta})\mathcal{N}(\mbf{x}_M|\mbf{0},\mbf{C}_\phi)\mathcal{N}(\mbf{y}|\mbf{x}_M,\sigma^2\mbf{I}) \mathcal{N}(\mbf{f}_M(\mbf{\theta},\mbf{x}_M,\tilde{\mbf{x}}_L(\mbf{x}_M,\mbf{\theta}))|\mbf{D}\mbf{x}_M,\mbf{A}+\gamma\mbf{I} ).
\end{align}
Finally, we get
\begin{align}
    & \rho(\mbf{x}_M,\mbf{\theta}|\mbf{y},\phi,\sigma,\gamma)\propto \notag\\
    & \rho(\mbf{\theta})\mathcal{N}(\mbf{x}_M|\mbf{0},\mbf{C}_\phi)\mathcal{N}(\mbf{y}|\mbf{x}_M,\sigma^2\mbf{I}) \mathcal{N}(\mbf{f}_M(\mbf{\theta},\mbf{x}_M,\tilde{\mbf{x}}_L(\mbf{x}_M,\mbf{\theta}))|\mbf{D}\mbf{x}_M,\mbf{A}+\gamma\mbf{I} ).
\end{align}
\end{proof}

\end{document}